# Smile for the Camera: Privacy and Policy Implications of Emotion AI

Elaine Sedenberg, John Chuang
School of Information
University of California, Berkeley
{elaine, chuang}@ischool.berkeley.edu

## Abstract

We are biologically programmed to publicly display emotions as social cues and involuntary physiological reflexes: grimaces of disgust alert others to poisonous food, pursed lips and furrowed brows warn of mounting aggression, and spontaneous smiles relay our joy and friendship. Though designed to be public under evolutionary pressure, these signals were only seen within a few feet of our compatriots — purposefully fleeting, fuzzy in definition, and rooted within the immediate and proximate social context.

The introduction of artificial intelligence (AI) on visual images for emotional analysis obliterates the natural subjectivity and contextual dependence of our facial displays. This technology may be easily deployed in numerous contexts by diverse actors for purposes ranging from nefarious to socially assistive — like proposed autism therapies. Emotion AI places itself as an algorithmic lens on our digital artifacts and real-time interactions, creating the illusion of a new, objective class of data: our emotional and mental states. Building upon a rich network of existing public photographs — as well as fresh feeds from surveillance footage or smart phone cameras — these emotion algorithms require no additional infrastructure or improvements on image quality.

Privacy and security implications stemming from the collection of emotional surveillance are unprecedented — especially when taken alongside other signals including physiological biosignals (e.g., heartrate or body temperature). Emotion AI also presents new methods to manipulate individuals by targeting political propaganda or fish for passwords based on micro-reactions. The lack of transparency or notice on these practices makes public inquiry unlikely, if not impossible.

In order to examine the potential policy and legal remedies for emotion AI as an emerging technology, we first establish a framework of actors, collection motivations, time scales, and space considerations that differentiates emotion AI from other algorithmic lenses. Each of these elements influences available policy remedies, and should shape continuing discussions on the antecedent conditions that make emotional AI acceptable or not in particular contexts.

Emotion analysis has great potential to add on to existing digital infrastructure with ease and result in a variety of benefits and risks. Based on our framework of unique elements, we examine potential available policy remedies to prevent or remediate harm. Specifically, our paper looks toward the regulatory role of the Federal Trade Commission in the US, gaps in the EU's General Data Protection Regulation (GDPR) allowing for emotion data collection, and precedent set by polygraph technologies in evidentiary and use restrictions set by law. We also examine the way social norms and adaptations could grow to also modulate broader use. Given the challenges in controlling the flow of these data, we call for further research and attention as emotion AI technology remains poised for adoption.

# I. Introduction

The allure of mind and mood reading applications has been an irresistible technical pursuit for decades—if not centuries—manifesting in technologies ranging from the polygraph to mood rings. Harnessing the ability to read the mental and emotional states of others (and perhaps even ourselves) gives us insight within numerous contexts, and has been a temptation of mystics and technologists alike. Despite this age-old quest to decipher our inner worlds, these signals ironically and almost literally sit right under our noses. Expressions and many physiological states were biologically designed to broadcast clues regarding our inner states. Yet even when within plain sight, signals are subtle, ambiguous, and simply complicated.

Our physiology programs us to publicly display emotions as social cues and involuntary reflexes: grimaces of disgust alert others to poisonous food, pursed lips and furrowed brows warn of mounting aggression, and spontaneous smiles relay our joy and friendship.[1] An entire spectrum of our emotions and many of our mental states (health, emotional and behavior patterns) are in essence a feed of open information scrolling across our most identifying feature: our faces. Though designed to be public under evolutionary pressure to be demonstrable, these signals were only seen within a few feet of our compatriots—purposefully fleeting, fuzzy in definition, and rooted within the immediate and proximate social context.[2] Only snippets of this open feed could be vaguely interpreted by others in close proximity.

The introduction of artificial intelligence (AI) on visual images for emotional analysis obliterates the natural subjectivity and contextual dependence of our facial displays.[3] New startups and applications build off of decades of research in affective computing: a field dedicated to the development of machines that can both recognize and display emotional responses during human computer interactions.[4] Suddenly emotion AI claims detect the visual difference between a genuine Duchenne smile reflex and a forced grin—but lack social debate on what it means to spontaneously erupt in glee verses an earnest but compelled smile.[5] This technology may be easily deployed in numerous contexts by diverse actors for purposes ranging from nefarious surveillance to socially assistive—like in proposed autism therapies.[6] Emotion AI places itself as an algorithmic lens on our digital artifacts and real-time interactions, creating the illusion of a new, objective class of data: our emotional and mental states. Building upon a rich network of existing public photographs—as well as fresh feeds from surveillance footage or smart phone cameras—these emotion algorithms require no additional infrastructure or improvements on image quality. In addition, visual analysis blends smoothly with other information clues like biosignals (e.g., heartrate or body temperature).

Privacy and security implications stemming from the collection of emotional surveillance are unprecedented—especially when taken alongside physiological signals or used alongside biometric authentication. Emotion AI also presents new methods to manipulate individuals by targeting political propaganda or fish for passwords based on micro-reactions, or simply make us vulnerable emotionally in ways never before imagined. The lack of transparency or notice on these practices makes public inquiry unlikely, if not impossible.

This paper first establishes background on the implicit relationship of biometrics, specifically facial recognition, to emotion analysis. We introduce studies from psychology and physiology on expressions and microexpressions, and other related literature on how our digital traces leave evidence of our mental health in order to fully explore the near-term potential of emotion analysis. We then establish a framework for examining motivations for use by unique actors, time dynamics, space considerations, and how these elements are both unique to emotion AI and influence policy options. We then outline potential policy remedies for privacy harms, and examine existing laws and case precedent for similar biometric and polygraph technologies. Our analysis points to gaps in the current privacy legal frameworks with regard to emotional data, and ends with a call for further research as this technology undergoes adoption from a range of actors.

## II. Background and Technical Foundations

In order to assess the social, legal, and policy implications of emotion AI, this paper will first establish current technological developments and boundaries. Biometrics refer to measurements of physical characteristics that can be used to identify an individual.[7] Biometrics include—but are not limited to—fingerprints, iris patterns, and particular facial features that have unique measurements between particular points that give some measure of statistical assurance of identity. Biometrics are not the same as biosensed information: data about heartrate, electrodermal activity (EDA), body temperature, electroencephalogram (EEG), etc. While biosensed information may be used to authenticate an identity, as exemplified by EEG-based authentication, data collected during everyday activities may be descriptive of our health, physiological state, behavior, and emotions yet are not enough to identify us uniquely.[8] Biosensed information may be sensed remotely with increasing resolution and commercial potential, and may be used to augment biometric data that links our identities to these descriptive (and often revealing) attributes.[9]

Facial recognition or identification is likely to go hand-in-hand alongside the adoption of emotional analytics since it shares a common and necessary data source: our face. In addition gait analysis and vocal recognition that might be possible in video, researchers in Japan have demonstrated that photos of someone exhibiting a peace sign from three meters away can be enough to recreate their fingerprints—enough to verify their identities or link to other biometric databases.[10] Our emotions are inextricably linked to our identities by the nature of how our bodies were designed.

Affective computing and facial emotion recognition technology is still under development. For instance, working with different camera angles or subject poses continues to present problems in accurate assessment.[11] Algorithms are trained on datasets collected by programmers and researchers (which poses challenges to produce ecologically valid emotion expressions), or on open repositories like the Affectiva-MIT Facial Expression Dataset (AM-FED).[12] Undoubtedly, the subjects and quality of training datasets will influence the extensibility and accuracy of emotion recognition algorithms developed by the private sector. There are also current technical developments to improve social signal analysis more broadly by including not just expressions, but body movements, physiological signals like heartrate, and vocal intonation.[13] Many of these signals could be picked up in video, or alongside other remote sensors to give a more complete and accurate picture.[14] For instance, studies have shown that moderate quality video data can be analyzed to reveal different blood flow patterns to the face—biological clues relating to our health, circulation, temperature, and sympathetic or parasympathetic stress responses.[15] Though this paper focuses mostly on expression analysis, emotion AI could be considered to include facial recognition linked to identity, as well as combination with these wider biosignals and contextual clues for a more intimate portrait of our emotional states.

Even if identities are not obtained during expression analysis, if underlying photos or video are storied retroactive identification would be possible. In some cases, our facial features may relay underlying genetic conditions like Down Syndrome, which has implications for inferences drawn about cognitive abilities. [16] Other genetic disorders, like mandibular prognathism—otherwise known as the "Habsburg Jaw"—has links to health as well as family associations due to its close link to incest and inbreeding. This particular genetic condition has been studied in medical literature through portraiture and written accounts of prominent families from 13th century to the introduction of photography.[17]

The ubiquity of cameras with increasing resolution creates infrastructure that make facial recognition and emotional analysis features easy additions to any surveillance system, camera-ready app, or collection of photographs. The iPhone8 is rumored to include facial recognition with infrared capabilities so that you can authenticate your device even when the room is dark, but it is also believed it can scan your facial features while lying flat on a table.[18] Additionally, the potential for tiny hidden cameras like those embedded in the head of a screw make surreptitious recording possible anywhere and at a resolution that would enable facial analytics.[19] There is no paucity of visual information already available to anyone with internet access—in 2014 the Internet Trends report estimated there were 1.8 billion digital images uploaded very day on various social media sites and repositories.[20]

Tech companies and researchers already mine existing data trails to attempt to understand our emotional states. For instance, one study by researchers established that the darker and more solo faces in Instagram posts may indicate whether or not you are experiencing depression.[21] One research group used language from Twitter (specifically angry words) to predict heart disease mortality at a community level.[22] If your selection of filters or public word choices may be an indicator of wellbeing, patterns of emotion in real life or posted images will certainly enable new 3rd parties to make inferences.

Our interpretation of others' emotional states in real life is flawed: colored by context and interpersonal history, our own emotional position, and even our own mental state.[23] Interestingly, previous studies have shown that there are few differences between cultures in identifying emotional expressions.[24] Other studies have shown that the intensity of emotions expressed and accompanying facial movements such as eye activity vary by culture, which would have bearing on the analysis and expectations that particular reactions correlate to particular ends.[25] Further our social exchanges, while nuanced, often involve complicated interdependencies with mimicry to reflect our conversation partner's mood in order to promote bonding and exhibit prosocial behavior.

Our microexpressions lie underneath outer facades or in fleeting looks, and betray concealed or subconscious emotional states that are most often too brief or subtle to notice in natural interactions.[26] Microexpressions from others may leave an impression after an interaction, but lack the certainty or explicit labeling of algorithmic verification—a shift that makes the subtle a virtual roar. Deception is a notoriously difficult expression and social cue to pick up on, but studies have shown that individuals who are able to acutely pick up on microexpressions were able to identifying lying behaviors.[27] Emotion AI with its ability to register microexpressions could in effect make the polygraph—as well as it's disputed integrity and scientifically credibility—available without the wires and expert operator.

When thinking about social behaviors, it is important to consider how the effect of "being watched" will manifest in social contexts once individuals are aware of emotion AI use. Research on privacy behaviors has shown that the presence of visible surveillance cameras has an impact on unruly public behavior.[28] Privacy literature also raises questions on how individuals will prefer to modulate their own privacy, as well as others around them. For instance, one study showed that participants with a camera wearable preferred real time physical control over information feeds rather than a burden of later review, and that participants were often very concerned about others' privacy around them even when others failed to voice any concerns.[29] These studies and countless others raise important questions for how behaviors could change in the known presence of emotion AI, and how individuals will choose to best modulate their own privacy, as well as the privacy of those around them.

## III. Framework of emotion AI differential elements

In order to examine the potential policy and legal remedies for emotion AI as an emerging technology, we first establish a framework of actors, collection motivations, time scales, and space considerations that differentiates emotion AI from other algorithmic lenses. Each of these elements influences available remedies, and should shape continuing discussions on the antecedent conditions that make emotional AI acceptable or not in particular contexts.

### Actors and Collection Motivations
How technology may be used and the ways society can use norms, laws, and policies to prevent harms or promote responsible uses differs greatly depending on the actor. Here we consider existing, imminent, or plausible uses with emotion AI.

*Governments*: Biometric technology has a long history of use by law enforcements to track individuals of interest and lawbreakers.[30] Now the FBI is able to compare mugshots (in addition to fingerprints) using the Interstate Photo Service (IPS). The FBI also relies on the FACE Services Unit and automated facial recognition during active investigations on photos collected during an investigation, and compares images to other government databases (including the State Department's Passport Photo File and Visa Photo File).[31] Facial recognition at border kiosks intended to identify individuals overstaying visas and general biometric collection procedures at visa stations are already in place.[32]

---

[28] Priks, Mikael. "Do Surveillance Cameras Affect Unruly Behavior? A Close Look at Grandstands." *Scandinavian Journal of Economics* 116, no. 4 (2014): 1160–79. doi:10.1111/sjoe.12075.

[29] Hoyle, Roberto, Robert Templeman, Steven Armes, Denise Anthony, David Crandall, and Apu Kapadia. "Privacy Behaviors of Lifeloggers Using Wearable Cameras." *UbiComp '14 - Proceedings of the 2014 ACM International Joint Conference on Pervasive and Ubiquitous*, 2014, 571–82. doi:10.1145/2632048.2632079.

[30] Prior to the widespread use of photography, physical descriptions were weak methods of tracking high interest individuals or repeat offenders. Fingerprints and other primitive biometrics were developed to assist in tracking across jurisdictions and time. (Nelson, Lisa S. *America Identified: Biometric Technology and Society*. The MIT Press, 2010.)

[31] Del Greco, Kimberly J. "Law Enforcement's Use of Facial Recognition Technology." Washington, D.C., 2017. https://www.fbi.gov/news/testimony/law-enforcements-use-of-facial-recognition-technology.

[32] Nixon, Ron. "Border Agents Test Facial Scans to Track Those Overstaying Visas." *New York Times*, August 1, 2017. https://www.nytimes.com/2017/08/01/us/politics/federal-border-agents-biometric-scanning-system-undocumented-immigrants.html?_r=0.; "Safety & Security of U.S. Borders: Biometrics." *U.S. Department of State Bureau of Consular Affairs*. Accessed August 10, 2017. https://travel.state.gov/content/visas/en/general/border-biometrics.html.

With this infrastructure in place, emotional screening could easily be deployed at the border, in crowds (e.g., protests), restricted or high security areas like airports or national monuments, during investigations of video/surveillance footage, and beyond. Automating emotional screening to identify emotions that put security at risk or indicate potential motives in a crime (anger, deceit, nervousness, alertness, etc.) could seemingly automate screening already done by trained border agents, TSA officers, or law enforcement. Of course, screening would come with the same ambiguities and risks of false positives/negatives that filtering by human eye and gut instinct does, except this time it would come with algorithmic validation. Additionally, emotional AI could be used by governments in high-stake negotiations, like in peace treaty negotiations, arms deals, and agreements of all kinds. Motivations to use emotion AI in government would likely stem from a desire to promote security, enforce laws, and decrease uncertainty in high-stakes decision making.

*Private Sector*: The private sector has direct monetary interests in developing and using emotion AI to sell products, improve or tailor experiences, or create apps and features that work with customer emotions in a way that engages and draws in users. Smart advertising that uses real-time tracking of attention (gaze detection) and emotional reactions has been demoed in some political advertising.[33] Additional close proof of concepts include smart shelf technology, which has been developed by Mondelez International—a snack food company—in order to customize in store experiences to shoppers. The technology, which is still under prototype, watches shopper's visual focus in addition to collecting and considering both age and gender demographics before creating a customized advertisement experience.[34] It would not take any additional sensors in order to augment smart shelves with emotion recognition and adaptation, and roll out the technology in any participating store.

Recently, Facebook has been under scrutiny for patents that enable computer and cell phone cameras to detect emotion of users while using the site so that platform content (peer-generated and advertising) can be tailored for the individual.[35] Even though the technology does not appear to be currently in use, it further demonstrates an appetite by the private sector to utilize emotion AI data.

In order to customize content and better serve advertising based on innate responses from customers, private companies are likely to learn more than just reactionary details about their users. For instance, behavioral patterns over time may relay mental states like depression, which has both ethical concerns as well as direct monetary payout if advertising for pharmaceutical drugs could be delivered. The tensions in being able to make money off of such customization will be deeply intertwined with our private "emotional baggage" and reveal more than intended to the private sector. Depending upon the permissions to use cameras on private computers and the associated privacy policy, or the legal presence of augmented cameras in public spaces, the use of emotion recognition technology is a simple addition to current platforms and spaces.

*Individual Use:* Emotion AI may be marketed and sold to consumers directly as apps, software, or built into different at home Internet of Things (IoT) devices and services. Individuals could use emotion AI as photo filters like those on Instagram or Snapchat to highlight particular emotions in photos shared with networks, or applications could easily run analysis in real time via phone cameras to offer seemingly instant emotional assessments. This could play out in a number of private contexts like on dates, parent to child, or during interpersonal moments of tension to establish a "ground truth". The motivations of emotion AI consumers would include, but is not limited to, truth seeking, exploration, social sharing, and self-reflection. It could also be used by individuals in more nefarious ways such as tricking and deceiving others by manipulating the algorithm, or secretly analyzing others' interactions during interpersonal interactions, etc.

Each of these actors bear different consequences but the same risk in misidentifying individuals, and misattributing particular emotions to individuals. Motivations and purposes of use range from public safety, profit, personal benefit or edification, and unknown application yet to be discovered.

**Time Dimensions**
An important element of emotion AI is that it can be run in three different modes of time: 1) Isolated retroactive; 2) Retroactive time series; and 3) Real-time analysis. Photos or videos from any point in time of a likely shrinking barrier of low resolution could be subject to emotion AI now or in the future. For instance, emotion analysis could be run on a repository of Facebook or Flickr photos at any time, potentially giving clues about past experiences and emotional states. These incidents would be isolated and without much contextual information about the events depending on the photo source. The presence of isolated retroactive analysis would be invisible to the photo subject, and when separated from information about the interaction, may easily be taken out of context. Similarly, retroactive time series might be photos that are all taken within a limited span of time (e.g., a birthday party or wedding) or on a video recording. There might be more contextual information available based on audio or the number of images surrounding interactions, and it may enable deeper analysis of mood changes over time. Both of these retroactive analytics could belie clues regarding our mental health or offer historical comparisons to our current moods.

Real-time analytics offer the ability to tailor experiences or interactions in real time. It may be that real-time analysis offers a chance for the subject to recognize they are on camera (e.g., if a smart phone is held up or it is obvious they are on camera), but even though it might be possible for the subject to know they are being recorded they would not have any indication they are under emotional scrutiny. It is also possible that real-time analytics are designed for automated personalization. Analysis may be conducted in real time, but differs in that the personalization might happen automatically based on feedback and data or analytical outputs are never reported back to a centralized repository.

 The time dimension of emotion AI and ability to be run in retrospection or in real-time has direct implications on potential remedies to mitigate harms and promote benefits.

**Space Considerations**
The location in which emotional analysis may be conducted might influence potential policies and laws. For instance, at sensitive security locations like borders and airports different rules and regulations may be put into place for government actors that might operate under unique search laws and surveillance norms. Public spaces like parks might be prime grounds for governments (local, state, federal) to monitor groups of people, or public spaces where individuals could bring technology to watch others in public where there is no reasonable expectation of privacy. Privately owned but otherwise public spaces like shopping centers may feature special pressures of property owners to emotionally surveil shoppers—even if technically legal—given interpretations of the space as public spaces where one would not be subject to scrutiny by a proprietor. Private property in physical space or perhaps online on particular platforms might be subject to different rules and expectations for emotional analysis, and might be ripe grounds for owners to utilize emotional analysis as they please. Further, locations like classrooms, hospitals, workplaces are likely to be highly contested spaces on the rights of individuals in each circumstance. A complete analysis of the privacy laws in each of these particular jurisdictions is beyond the scope of this paper. However, it is important to consider the role "space" may play when deploying emotion AI and how different public or private spaces evoke different search regulations, privacy expectations, and social norms.

**Enhancing Factors**
The ability to combine emotional with other data streams like audio for tonal analysis and social context, patterns over time (especially long periods), and other biosensed data like heartrate, body temperature, electrodermal activity, and others could enhance facial expression capture. For instance, physiological signs of stress like increased heartrate, elevated body temperature, and presence of skin sweat could augment emotional analysis of worry by giving clues to the intensity of the emotion felt. Supplemental information could also help disambiguate otherwise confusing emotions by giving clues whether emotions stem from physical stress or mental anguish. It is unlikely emotion AI will develop in isolation from these other sensing modalities. As discussed in the previous section, some of these addition elements may be available in moderate resolution photographs, or could be added with additional sensors to real-time analysis. When considering remedies in the next section, it will be important to consider the role these enhancing information streams (mostly subject to the same current legal and regulatory frameworks) could play in making the results more invasive and sensitive.

# IV. Controls, Remedies, and Policy Considerations

Controlling the use of digital technology—and in particular, algorithmically based tech—is a challenge for policymakers aiming to balance to benefits and risks for society. By looking at how privacy laws and regulations are currently structured, we examine how current infrastructure in the US and EU may regulate emotion analysis use. We then examine how social norms and behavioral adaptations may influence the use of this technology.

**Legal and policy constraints and remedies**

In the United States, the Federal Trade Commission (FTC) has become the defacto regulator of privacy by way of its consumer protection mission. Through the FTC's unfair and deceptive trade practice authority, the agency would be able to regulate consumer products that make emotion AI available to consumers or utilize these features in existing services. Such a product could result in unjustified consumer injury to a consumer's mental wellbeing, or opening up discrimination by labeling someone as "mentally ill." Additionally, services could feasibly deceive consumers through "a representation, omission, or practice that is likely to mislead" the average consumer when advertising or touting the benefits of emotion AI products.[36] Alternatively, emotion analysis could be baked into other consumer services in a way that violates their terms of service or harms and/or

---

deceives the average consumer. One could imagine a platform running emotional AI on photographs posted under different terms and expectations, and using that information to tailor services that could result in harm or public exposure to those with a mental health issue. The FTC would only be a remedy to harms in by the private sector in the consumer products and services, but would not address other actors like governments.

Broader regulation of emotional information as a class of data—similar to how health data is protected and regulated in the US—could be feasibly possible if clear definitions could be drawn on what constitutes "personal emotion data," given that inferences and patterns gleaned are the more important than the data itself. The EU General Data Protection Regulation (GDPR) specifically names "biometric data" as a special category of personal data that cannot be processed without first satisfying established permissions or exceptions.[37] However, the regulations leave emotion tracking unregulated so long as the emotion analytics do not allow or confirm the unique identification of an individual. Since emotion analytics are often tied to faces or even voice (both of which are obviously linked to identity) the GDPR prevents the recording of these sentiments and tracking with an identity. Yet, some harms stemming from emotion AI could be invisibly adapting advertising or customization, which do not require storage of images or linkage to identity.[38] Also, data about general demographics (age, gender, race, etc.) could be linked in a way so that groups of people are generalized together, or that groups of people in a crowd could be assessed for an intervention (e.g., an assessment of a protest determining whether or not they should be allowed to continue the assembly). As privacy laws are developed and revised, there should be a consideration for biosensed data that are not considered biometrics, as well as considerations for issues of group privacy.

The treatment of lie detector technology offers a unique model for handling result ambiguity and potential disagreements between experts on the veracity and reliability of technology. The first lie detector based upon systolic blood pressure was developed by William Marston in 1917, and in refused as evidence in a 1923 court case after the court of appeals determined that "the systolic blood pressure deception test has not yet gained such standing and scientific recognition among physiological and psychological authorities as would justify the courts in admitting expert deduced from the discovery, development, and experiments thus far made."[39] After this foundational case in evidence law, the modern polygraph machine was developed to collect three separate physiological responses including electrodermal activity, relative blood pressure, and respiration rate—yet these changes have not altered the status of polygraphs as admissible evidence in many courts since it did not address concerns by some experts in how the machine and operator reaches an assessment. The Supreme Court affirmed the lack of scientific evidence and expert agreement in a 1998 case stating the use of the polygraph "is no more accurate than a coin flip."[40] Polygraphs are not permitted as evidence in most courts, but may be submitted in certain states on particular rules.[41] The use of polygraphs in workplace settings was regulated by the Employee Polygraph Protection Act (EPPA) in 1988 following several court cases.[42] EPPA places limitations on polygraph use in the private sector, but exempts federal, state and local governments.

It is likely any emotional AI will be contested by experts for at least some period of time, and it is plausible that it will follow similar evidential and use restrictions in particular contexts (e.g., the workplace) as it is litigated over time. Polygraphs offer a model for how lawmakers might be able to limit harmful outcomes in an enforceable way of a technology with contested results and reliability.

**Social norms and behavioral adaptations**
In the case of many socially disruptive technologies, people evolve social norms or behavioral adaptations to address some risks and harms associated with use. For instance, even though it is legal for someone to walk around taking photos on their iPhone of others in public or in one-on-one settings, it is socially taboo in many contexts. If one does not want their photograph taken, they can hold up their hand or turn the other way. Even if an individual attempts to obscure their photography, it still needs to be within line of sight in order to get the shot and usually sends a behavioral signal unless it is a hidden camera. This matters in the cases of emotional AI use because it will rely on a visual information feed, but the analysis portion is invisible to any potential subject. This complicates the ability to obfuscate the line of sight or adapt social norms given this hidden algorithmic component, however it could still become taboo to run emotional analysis retroactively on someone—somewhat how Facebook stalking became an activity done in private without discussion, and maybe something as the platform ages people skip altogether.

There may be other social adaptations that allow individuals to prevent the use of emotion AI on them, like the use of make-up designed to trick the algorithms or fashion adaptations like veils, large sunglasses, or even flash resistant cloth used in scarves or other garments.[43] Even in the absence of legal remedies or in cases where enforcement is a challenge, there may be other ways to socially regulate and steer use toward socially accepted practices.

# V. Conclusion

Emotion analysis has great potential to add on to existing infrastructure with ease and result in a variety of benefits and risks. Our research points to the unique privacy and social implications of emotion AI technology, and the impact it may have on both communities and individuals. Based on our framework of unique elements based upon actor, collection motivation, time, and space, we examine potential available policy remedies to prevent harm and promote beneficial uses. Looking toward analogous privacy laws and regulations, we illustrate the ways emotional AI could be regulated within the EU and US laws, or the way social norms and adaptations could grow to also modulate use. Given the challenges in controlling the flow of these data, we call for further research and attention as emotion AI technology remains poised for adoption.

---

[43] See, for example: ACCESSALLBRANDS. "ISHU." Accessed July 1, 2016. https://theishu.com/.

# VI. Acknowledgements

This work was funded and made possible by a generous grant from the Hewlett Foundation and Center for Long-Term Cybersecurity (CLTC) at the University of California, Berkeley.

This material is also based upon work supported by the National Science Foundation Graduate Research Fellowship Program under Grant No. DGE1106400. Any opinions, findings, and conclusions or recommendations expressed in this material are those of the author(s) and do not necessarily reflect the views of the National Science Foundation.

This manuscript was prepared for and presented at TPRC45 in September 2017.